\documentclass[aps,prb,twocolumn,superscriptaddress]{revtex4}
\usepackage{graphicx}
\usepackage{bm}
\usepackage{amsmath,amssymb,amsfonts,bm}
\usepackage{epsfig}
\usepackage{epstopdf}
\usepackage{dcolumn}
\usepackage{grffile}
\usepackage{verbatim}
\usepackage{mathrsfs}
\usepackage[colorlinks=true,linkcolor=blue,citecolor=blue, urlcolor=blue]{hyperref}

\begin{document}


\title{Exact Solution to the Haldane-BCS-Hubbard Model Along the Symmetric Lines: Interaction Induced Topological Phase Transition}

\author{Jian-Jian Miao}
\affiliation{Kavli Institute for Theoretical Sciences and CAS Center for Excellence in Topological Quantum Computation, University of Chinese Academy of Sciences, Beijing 100190, China}

\author{Dong-Hui Xu}
\affiliation{Department of Physics, Hubei University, Wuhan 430062, China}

\author{Long Zhang}
\affiliation{Kavli Institute for Theoretical Sciences and CAS Center for Excellence in Topological Quantum Computation, University of Chinese Academy of Sciences, Beijing 100190, China}
\affiliation{Physical Science Laboratory, Huairou National Comprehensive Science Center, Beijing 101400, China}

\author{Fu-Chun Zhang}
\affiliation{Kavli Institute for Theoretical Sciences and CAS Center for Excellence in Topological Quantum Computation, University of Chinese Academy of Sciences, Beijing 100190, China}
\affiliation{Physical Science Laboratory, Huairou National Comprehensive Science Center, Beijing 101400, China}

\date{\today}

\begin{abstract}

We propose a Haldane-BCS-Hubbard model on a honeycomb lattice, which is composed of two copies of the Haldane model of the quantum anomalous Hall effect, an equal-spin pairing term and an onsite Hubbard interaction term. 
For any interaction strength, this model is exactly solvable along the symmetric line where the hopping and pairing amplitudes are equal to each other. 
The ground state of the Haldane-BCS-Hubbard model is a topological superconducting state at weak interaction with two chiral Majorana edge states. A strong interaction drives the system across a topological quantum phase transition to a topologically trivial superconductor. A $\mathbb{Z}_{2}$ symmetry of the Hamiltonian, which is a composition of the bond-centered inversion and a gauge transformation, is spontaneously broken by the interaction, resulting a finite antiferromagnetic order in the $y$-direction.

\end{abstract}

\maketitle


\section{Introduction}

The concept of topology in condensed matter physics has flourished in the past decades\cite{Hasan_RMP,Qi_RMP}.
This abstract notion is deeply related to the band structure in the momentum space.
The topological band theory has been established and a lot of predicted materials have been synthesized\cite{Bansil_RMP}.
Recently the full diagnosis of the non-trivial band topology for non-magnetic materials have been established\cite{Fang,Bernevig,Wan}.

The interplay of topology and correlations can lead to novel phases and phase transitions in condensed matter systems.
First, the interactions may reduce the topological classification of free fermions in one dimension\cite{Fidkowski_2010,Fidkowski_2011} and two dimensions\cite{Yao}.
Second, interactions may drive topological quantum phase transitions, which is demonstrated in exactly solvable models of interacting Kitaev chains \cite{Miao,Ezawa,Wang}, the Haldane-Hubbard model \cite{Zhai} and the $\mathbb{Z}_{2}$ Bose-Hubbard model \cite{Daniel}.
Recently, Chen et. al.\cite{Ng_2018} generalized the construction of the Kitaev honeycomb model\cite{Kitaev} to spinful fermion models with both equal-spin pairing and Hubbard interaction terms, dubbed BCS-Hubbard model, which can be solved exactly when the pairing amplitude equals the hopping amplitude. 
Later Ezawa\cite{Ezawa_2018} generalized the BCS-Hubbard model on a honeycomb lattice by introducing the Kane-Mele spin-orbit coupling (SOC). 
However, an infinitesimal Hubbard interaction $U$ will destroy the topological superconducting state due to the spontaneous time reversal symmetry breaking in Ref. \onlinecite{Ezawa_2018}.
It is still desirable to find an exactly solvable model in two dimensions with topological phase transition at finite interaction strength to study the interplay of topology and correlations.

In this paper, we investigate the Haldane-BCS-Hubbard model on a honeycomb lattice.
Along the symmetric lines where the hopping amplitude equals the pairing amplitude, the model is exactly solvable and reduces to the Falicov-Kimball model\cite{Falicov}.
There is an interaction induced topological phase transition at finite Hubbard $U$ along the symmetric lines. The phase transition can be characterized by the change of the spectral Chern number.
Thus the topological superconducting state in our model is stable to small interaction.
These results are obtained exactly without approximation, and can serve as a benchmark for further study.

The paper is organized as follows:
In section \ref{model}, we introduce the Haldane-BCS-Hubbard model.
Then we show the exact solvability of the model along the symmetric lines in section \ref{exact}.
We analyze the symmetry of the model in section \ref{symmetry} and introduce the composite fermion representation in section \ref{composite} for later convenience.
In section \ref{noninteracting}, we study the noninteracting limit of the model and give the phase diagram.
In section \ref{interacting}, we study the model along the symmetric lines and show the interaction induced topological phase transition.
We summarize the results and propose the possible realization of the model in section \ref{discussion}.

\begin{figure}[b]
\begin{center}
\includegraphics[width=0.45\textwidth]{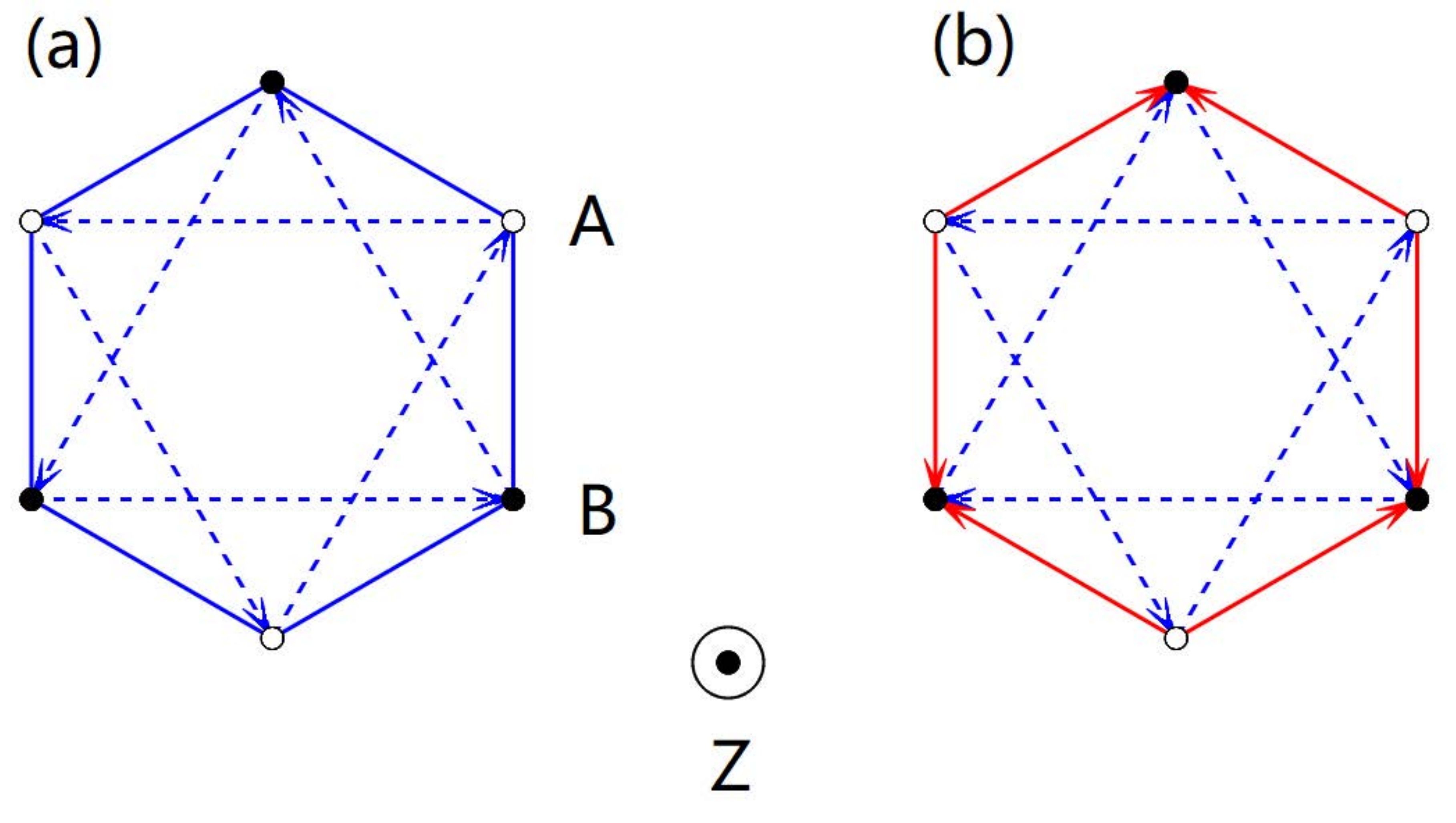}
\end{center}
\caption{(Color online) (a) The signs $\nu_{ij}$ of the spin-orbit coupling terms.  Along the direction of the dashed lines $\nu_{ij}=+1$, while along the reversed direction $\nu_{ij}=-1$. (b) The signs of the nearest  and next-nearest neighbor pairing terms.  Along the direction of the red (dashed) lines the signs of $\Delta_{1}$ ($\Delta_{2}$) are $+1$, while along the reversed direction they are $-1$. White sites $A$ and black sites $B$ are the two sublattices. $z$-direction points outward the plane.}
\label{fig:sign}
\end{figure}

\section{Model Hamiltonian}\label{model}
In this section, we introduce the Haldane-BCS-Hubbard model we study.
The Hamiltonian of the model consists of three parts and can be expressed as follows
\begin{equation}
\label{eq:ham}
H=H_{\mathrm{hop}}+H_{\mathrm{pair}}+H_{\mathrm{int}}
\end{equation}
where $H_{\mathrm{hop}}$ describes the electron hopping terms, which is a spinful generalization of the Haldane model\cite{Haldane,Sheng}, $H_{\mathrm{pair}}$ describes the equal spin pairing (ESP) terms, and $H_{\mathrm{int}}$ describes the on-site Hubbard interaction. They are given by
\begin{align}
H_{\mathrm{hop}} & =t_{1}\sum_{\left\langle ij\right\rangle s}c_{is}^{\dagger}c_{js}-t_{2}\sum_{\left\langle \left\langle ij\right\rangle \right\rangle s}i\nu_{ij}c_{is}^{\dagger}c_{js}\\
H_{\mathrm{pair}} & =\Delta_{1}\sum_{\left\langle ij\right\rangle s}c_{is}^{\dagger}c_{js}^{\dagger}+\Delta_{2}\sum_{\left\langle \left\langle ij\right\rangle \right\rangle s}i\nu_{ij}\lambda_{i}c_{is}^{\dagger}c_{js}^{\dagger}+h.c.\\
H_{\mathrm{int}} & =U\sum_{i}\bigl(n_{i\uparrow}-\frac{1}{2}\bigr)\bigl(n_{i\downarrow}-\frac{1}{2}\bigr)
\end{align}
where $c_{is}$ ($c_{is}^{\dagger}$) annihilates (creates) an electron at site $i$ with spin $s=\uparrow,\downarrow$ pointing in the $z$-direction. $\left\langle ij\right\rangle $ and $\left\langle \left\langle ij\right\rangle \right\rangle $ denote the nearest and next-nearest neighbor sites.
$t_1$ is the nearest neighbor hopping matrix element and $t_2$ is the next-nearest neighbor hopping (NNNH) matrix.
The sign $\nu_{ij}=\text{sign}\bigl(\hat{d}_{1}\times\hat{d}_{2}\bigr)_{z}=\pm1$,
where $\hat{d}_{1,2}$ are the vectors along the two bonds constituting the next-nearest neighbors.
The signs $\nu_{ij}$ are shown in FIG. \ref{fig:sign}.
$\Delta_1$ and $\Delta_2$ are the nearest and next-nearest neighbor ESP potential respectively. $\lambda_i=\pm1$ for sublattices $A$ and $B$ respectively. The signs of $\Delta_{1,2}$ are shown in FIG. \ref{fig:sign}. 
$n_{is}=c_{is}^{\dagger}c_{is}$ is the fermionic number operator for spin $s$.
$U$ is the strength of on-site Hubbard interaction.

\section{Exact solvability}\label{exact}
In this section, we shall show the exact solvability of the Haldane-BCS-Hubbard model along the symmetric lines 
\begin{equation}
\label{eq:sym}
t_1=\Delta_1, \quad t_2=\Delta_2.
\end{equation}
The Haldane-BCS-Hubbard model is not exactly sovable in general.  However, similar to the BCS-Hubbard model\cite{Ng_2018}, we find this model can be solved exactly along the symmetric lines.
The exact solvability of the model becomes manifest in the Majorana fermion representation.  
As the system contains two sublattices, we use $r$ to denote the unit cell and $c_{rs\lambda}$ to denote the annihilation operator with spin $s$ at unit cell $r$ in sublattice $\lambda=A,B$.
We then decompose the complex fermion operators $c_{rs\lambda}$ into Majorana fermion operators $\eta_{rs\lambda}$ and $\gamma_{rs\lambda}$ as follows
\begin{align}
c_{rsA} & =\eta_{rsA}+i\gamma_{rsA}\nonumber \\
c_{rsB} & =\gamma_{rsB}+i\eta_{rsB}
\end{align}
Note the decomposition is opposite for the two sublattices.
The Hamiltonian in the Majorana fermion representation becomes
\begin{widetext}
\begin{align}
H_{0} & =\delta_{1}\sum_{rs}\left(i\eta_{rsA}\eta_{rsB}+i\eta_{rsA}\eta_{r+a_{1}sB}+i\eta_{rsA}\eta_{r+a_{2}sB}\right)\nonumber \\
 & -\tilde{t}_{1}\sum_{rs}\left(i\gamma_{rsA}\gamma_{rsB}+i\gamma_{rsA}\gamma_{r+a_{1}sB}+i\gamma_{rsA}\gamma_{r+a_{2}sB}\right)\nonumber \\
 & -\delta_{2}\sum_{rs\lambda}\lambda\left(i\eta_{rs\lambda}\eta_{r+a_{1}s\lambda}+i\eta_{rs\lambda}\eta_{r-a_{1}+a_{2}s\lambda}+i\eta_{rs\lambda}\eta_{r-a_{2}s\lambda}\right)\nonumber \\
 & -\tilde{t}_{2}\sum_{rs\lambda}\lambda\left(i\gamma_{rs\lambda}\gamma_{r+a_{1}s\lambda}+i\gamma_{rs\lambda}\gamma_{r-a_{1}+a_{2}s\lambda}+i\gamma_{rs\lambda}\gamma_{r-a_{2}s\lambda}\right)\nonumber \\
H_{\mathrm{int}} & =U\sum_{r\lambda}2i\eta_{r\uparrow\lambda}\gamma_{r\uparrow\lambda}2i\eta_{r\downarrow\lambda}\gamma_{r\downarrow\lambda}
\end{align}
\end{widetext}
where $H_{0} =H_{\mathrm{hop}}+H_{\mathrm{pair}} $ and $\delta_{1}=2\left(t_{1}-\Delta_{1}\right)$, $\delta_{2}=2\left(t_{2}-\Delta_{2}\right)$,
$\tilde{t}_{1}=2\left(t_{1}+\Delta_{1}\right)$, $\tilde{t}_{2}=2\left(t_{2}+\Delta_{2}\right)$.
$a_{1}=a\left(\frac{1}{2},\frac{\sqrt{3}}{2}\right)$ and $a_{2}=a\left(-\frac{1}{2},\frac{\sqrt{3}}{2}\right)$ are the two basis vectors.
Note the $\eta$ Majorana fermions disappear in the noninteracting Hamiltonian $H_0$ along the symmetric lines $\delta_1=\delta_2=0$.  
In the following, we shall focus on the symmetric line. We define $\hat{D}_{r\lambda}=4i\eta_{r\uparrow\lambda}\eta_{r\downarrow\lambda} $.
It is easy to prove that $\left[\hat{D}_{r\lambda},H\right]=0$. Thus $\hat{D}_{r\lambda}$ are constants of motion. Since $\hat{D}_{r\lambda}^{2}=1$, we can replace the operators $\hat{D}_{r\lambda}$ by its eigenvalues $D_{r\lambda}=\pm1$. The Hubbard interaction becomes
\begin{equation}\label{Hint_Majorana}
H_{\mathrm{int}}=-U\sum_{r\lambda}D_{r\lambda}i\gamma_{r\uparrow\lambda}\gamma_{r\downarrow\lambda}
\end{equation}
The total Hilbert space is divided into different sectors characterized by $\left\{ D_{r\lambda}\right\} $. Within each sector, the Hamiltonian contains only quadratic terms of $\gamma$ Majorana fermions and can be solved exactly.

\section{symmetry}\label{symmetry}
Symmetry plays an important role in the following analysis. 
In this section, we shall analyze the various symmetry of the Haldane-BCS-Hubbard model.
The fermion operators transform as $Cc_{is}C^{-1}=\lambda_{i}c_{is}^{\dagger}$ under the particle-hole symmetry (PHS). It is obvious that the Hamiltonian has the PHS\cite{Chiu}.
The time reversal symmetry (TRS) operator for spinful system is $T=i\sigma^{y}K$, where $K$ denotes the complex conjugation and $T^{2}=-1$. 
The fermion operators transform as $Tc_{is}T^{-1}=\sum_{s'}\left(i\sigma^{y}\right)_{ss'}c_{is'}$ under TRS.
Just as in the Haldane model, the NNNH terms break the TRS explicitly. 
The sublattice symmetry (SLS) can be implemented by the bond centered inversion operator $I$. 
The signs shown in FIG. \ref{fig:sign} indicate $H_{\mathrm{pair}}$ breaks the SLS.
With NNNH and ESP terms, the Hamiltonian does not preserve the $SU\left(2\right)$ spin rotation symmetry.
The $SU\left(2\right)$ symmetry is reduced to $U\left(1\right)_{y}\rtimes \mathbb{Z}_{2}$, where $U\left(1\right)_{y}$ is the rotation about the $y$-axis and $\mathbb{Z}_{2}$ is the $\pi$-rotation around the $z$-axis.
Therefore, the system falls into class $D$ in the topological classification of superconductors (SC) \cite{Schnyder}. 

\section{composite fermion representation}\label{composite}
In this section, we introduce the composite fermion representation. These composite fermions form the quasiparticles for the Haldane-BCS-Hubbard model.
We define the composite fermions as in Ref. \onlinecite{Ng_2018}
\begin{align}
d_{r2\lambda} & =\eta_{r\uparrow\lambda}+i\lambda\eta_{r\downarrow\lambda}\nonumber \\
d_{r1\lambda} & =\gamma_{r\uparrow\lambda}-i\lambda\gamma_{r\downarrow\lambda}
\end{align}
The physical meaning of the composite fermions becomes clear by introducing the fermion operators pointing in the $\pm y$-direction as follows
\begin{equation}
c_{r\pm\lambda}^{\dagger}=\frac{1}{\sqrt{2}}\left(c_{r\uparrow\lambda}^{\dagger}\pm ic_{r\downarrow\lambda}^{\dagger}\right)
\end{equation}
We express $d$ composite fermions in terms of $c$ fermion operators
\begin{align}
d_{r2A}=\frac{c_{r-A}+c_{r+A}^{\dagger}}{\sqrt{2}},\quad & d_{r2B}=\frac{c_{r+B}-c_{r-B}^{\dagger}}{\sqrt{2}i}\nonumber \\
d_{r1A}=\frac{c_{r+A}-c_{r-A}^{\dagger}}{\sqrt{2}i},\quad &d_{r1B}=\frac{c_{r-B}+c_{r+B}^{\dagger}}{\sqrt{2}}
\end{align}
which are equal-weight superposition of particle and hole of $c$ fermion operators. $d_{r1A}$ and $d_{r2B}$ ($d_{r2A}$ and $d_{r1B}$) carry spin-$1/2$ pointing in the $y$($-y$)-direction. 
We write the Hamiltonian in the composite fermion representation
\begin{widetext}
\begin{align}
H_{0} & =\frac{i\delta_{1}}{2}\sum_{r}\left(d_{r2A}^{\dagger}d_{r2B}^{\dagger}+d_{r2A}^{\dagger}d_{r+a_{1}2B}^{\dagger}+d_{r2A}^{\dagger}d_{r+a_{2}2B}^{\dagger}\right)\nonumber \\
 & -\frac{i\tilde{t}_{1}}{2}\sum_{r}\left(d_{r1A}^{\dagger}d_{r1B}^{\dagger}+d_{r1A}^{\dagger}d_{r+a_{1}1B}^{\dagger}+d_{r1A}^{\dagger}d_{r+a_{2}1B}^{\dagger}\right)\nonumber \\
 & -\frac{i\delta_{2}}{2}\sum_{r\lambda}\lambda\left(d_{r2\lambda}^{\dagger}d_{r+a_{1}2\lambda}+d_{r2\lambda}^{\dagger}d_{r-a_{1}+a_{2}2\lambda}+d_{r2\lambda}^{\dagger}d_{r-a_{2}2\lambda}\right)\nonumber \\
 & -\frac{i\tilde{t}_{2}}{2}\sum_{r\lambda}\lambda\left(d_{r1\lambda}^{\dagger}d_{r+a_{1}1\lambda}+d_{r1\lambda}^{\dagger}d_{r-a_{1}+a_{2}1\lambda}+d_{rs\lambda}^{\dagger}d_{r-a_{2}1\lambda}\right)\nonumber \\
 & +h.c.\nonumber \\
H_{\mathrm{int}} & =U\sum_{r\lambda}\left(n_{r2\lambda}-\frac{1}{2}\right)\left(n_{r1\lambda}-\frac{1}{2}\right)
\end{align}
\end{widetext}
where $n_{r\alpha\lambda}=d_{r\alpha\lambda}^{\dagger}d_{r\alpha\lambda}$ with $\alpha=1,2$.
Thus the original system can be viewed as two species of $d$ composite fermions with nearest neighbor pairing, next-nearest neighbor hopping, and they interact with on-site Hubbard $U$.
Note the Hamiltonian $H$ has the dual symmetry under the dual mapping $d_{r1\lambda}\leftrightarrow d_{r2\lambda}$ (or $\eta_{rs\lambda}\leftrightarrow\gamma_{rs\lambda}$), with parameters changing as $\delta_{1}\leftrightarrow-\tilde{t}_{1}$, $\delta_{2}\leftrightarrow\tilde{t}_{2}$.
Thus the Hamiltonian $H$ has a self-dual point $t_1=\Delta_2=0$, even with the Hubbard interaction $U$.
In the following, we shall analyze the properties of the Haldane-BCS-Hubbard model in terms of $d$ composite fermions.

\begin{figure}[tb]
\begin{center}
\includegraphics[width=10cm]{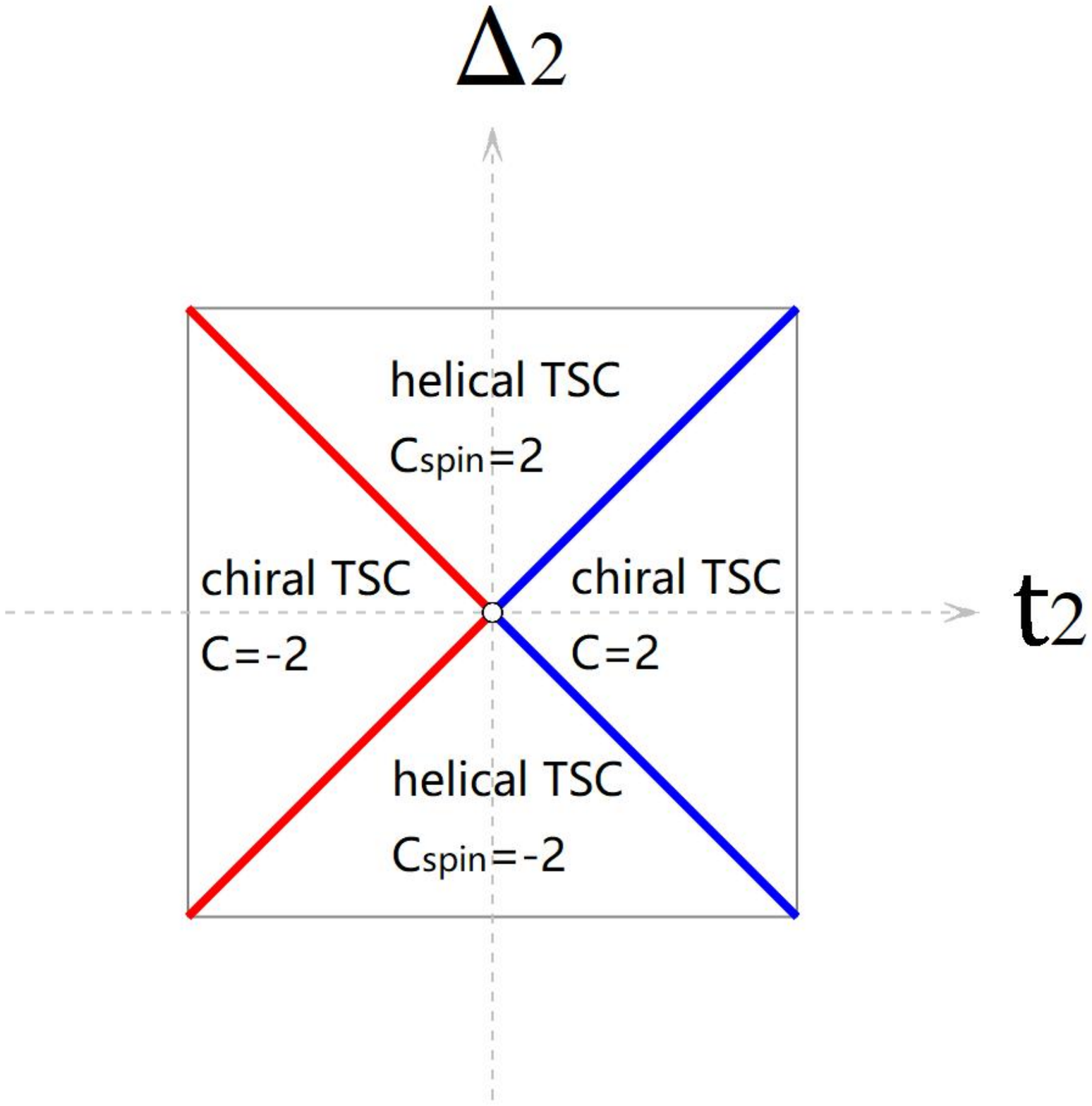}
\end{center}
\caption{(Color online) The phase diagram of the Haldane-BCS model with $t_{1}\neq\Delta_{1}$. Chiral TSC denotes the chiral topological superconducting state with total Chern number $C=\pm2$ and spin Chern number $C_{\mathrm{spin}}=0$. Helical TSC denotes the helical topological superconducting state with total Chern number $C=0$ and spin Chern number $C_{\mathrm{spin}}=\pm2$. The blue (red) lines denote one species of $d$ composite fermions is gapless and another species is in chiral TSC with Chern number $C=1$ ($C=-1$). The origin is a multicritical point.}
\label{fig:phase}
\end{figure}

\section{Noninteracting limit: Haldane-BCS model}\label{noninteracting}
In this section, we analyze the noninteracting limit of the Haldane-BCS-Hubbard model.
At $U=0$, the model reduces to the Haldane-BCS model.
Note the Hamiltonian $H_0$ is decoupled for two species of $d$ composite fermions $H_{0}=H_{1}+H_{2}$, where $H_{\alpha}$ contains $d_{\alpha}$ composite fermions only. 
The Hamiltonian $H_0$ is uniform and we can perform the Fourier transformation to obtain the spectrum.
The Fourier transformation is defined as
\begin{equation}
d_{r\alpha\lambda}=\frac{1}{\sqrt{N}}\sum_{k}e^{ik\cdot r}d_{k\alpha\lambda}
\end{equation}
We also define the spinor as $\psi_{k\alpha}^{\dagger}=\bigl(d_{k\alpha A}^{\dagger},d_{-k\alpha B}\bigr)$, the Hamiltonian $H_\alpha$ can be written in the form of 
\begin{equation}
H_{\alpha}=\sum_{k}\psi_{k\alpha}^{\dagger}h_{\alpha}\left(k\right)\psi_{k\alpha}
\end{equation}
where
\begin{equation}
h_{\alpha}\left(k\right)=\vec{T}_{\alpha}\left(k\right)\cdot\vec{\sigma}
\end{equation}
$\vec{\sigma}=\left(\sigma^{x},\sigma^{y},\sigma^{z}\right)$ are the Pauli matrices and $\vec{T}_{1}\left(k\right)$ are given by
\begin{align}
T_{1}^{x}\left(k\right) & =\frac{\tilde{t}_{1}}{2}\left(\sin k\cdot e_{1}+\sin k\cdot e_{2}+\sin k\cdot e_{3}\right)\nonumber \\
T_{1}^{y}\left(k\right) & =\frac{\tilde{t}_{1}}{2}\left(\cos k\cdot e_{1}+\cos k\cdot e_{2}+\cos k\cdot e_{3}\right)\nonumber \\
T_{1}^{z}\left(k\right) & =\tilde{t}_{2}\left(\sin k\cdot a_{1}-\sin k\cdot\left(a_{1}-a_{2}\right)-\sin k\cdot a_{2}\right)
\end{align}
where $e_{1}=a\left(0,-\frac{1}{\sqrt{3}}\right)$, $e_{2}=a\left(\frac{1}{2},\frac{1}{2\sqrt{3}}\right)$ and $e_{3}=a\left(-\frac{1}{2},\frac{1}{2\sqrt{3}}\right)$ are the three vectors of nearest neighbor bonds. 
$\vec{T}_{2}\left(k\right)$  can be obtain by the dual mapping $d_{r1\lambda}\rightarrow d_{r2\lambda}$ with parameters changing as $\tilde{t}_{1}\rightarrow-\delta_{1}$ and $\tilde{t}_{2}\rightarrow\delta_{2}$. 
The energy dispersions read $E_{\alpha}\left(k\right)=\pm\bigl|\vec{T}_{\alpha}\left(k\right)\bigr|$, which form reflects the PHS of the Hamiltonian. 
The ground state is unique with all the negative energy levels of both $d$ composite fermions are occupied. The system is gapped for nonzero $\tilde{t}_{1},\tilde{t}_{2}$ and $\delta_{1},\delta_{2}$.

\begin{figure*}[tb]
\begin{center}
\includegraphics[width=6.5cm]{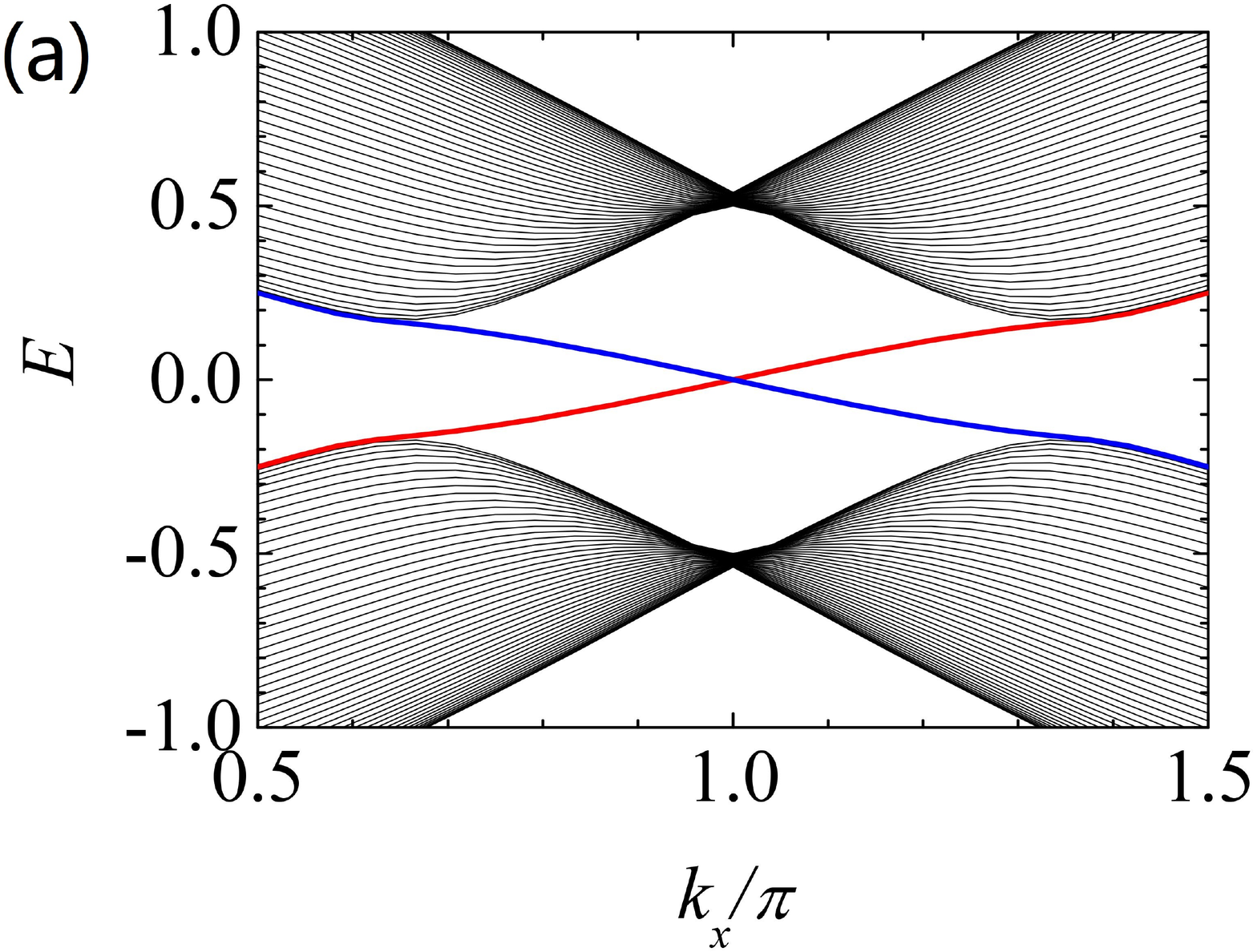}
\includegraphics[width=6.5cm]{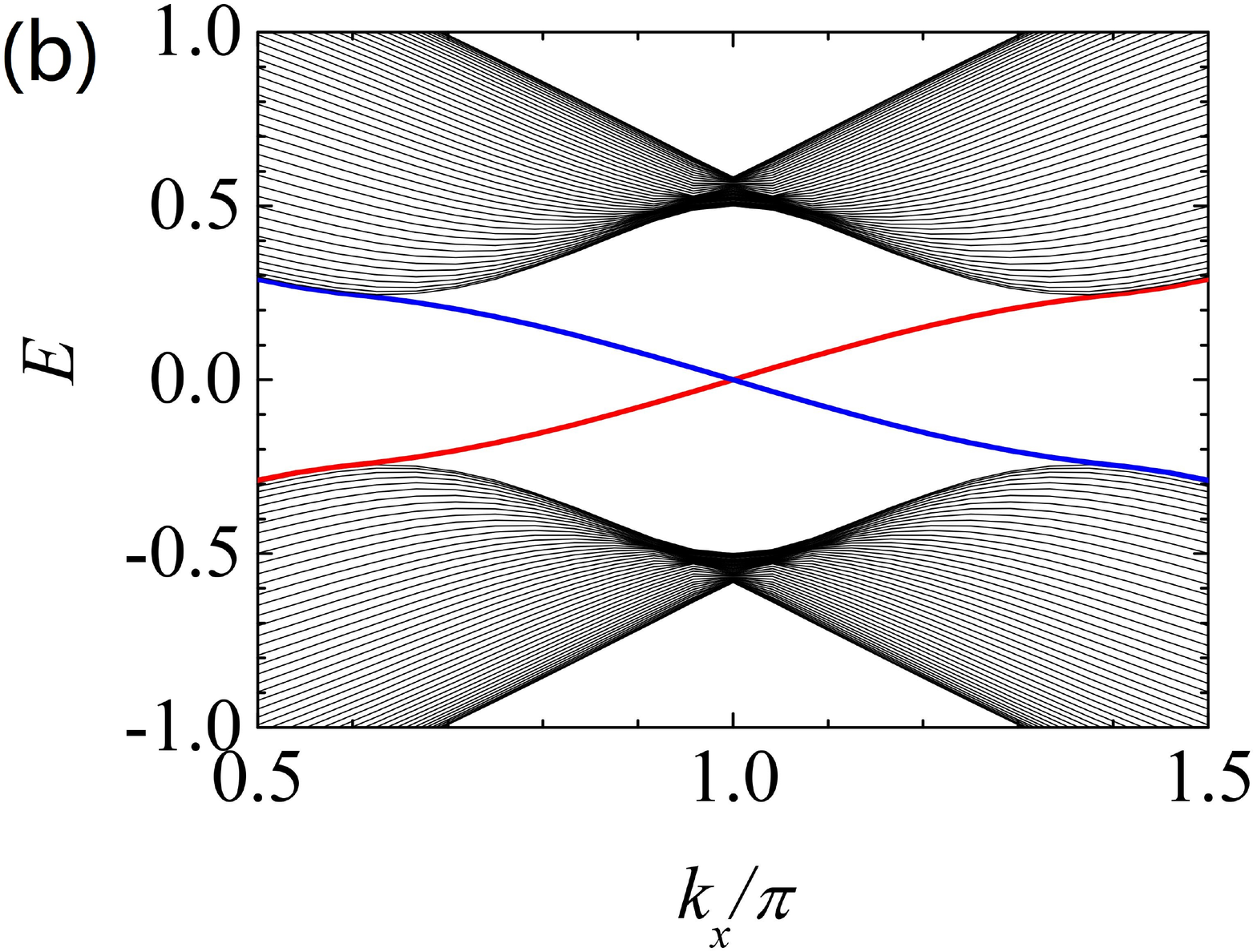}
\includegraphics[width=6.5cm]{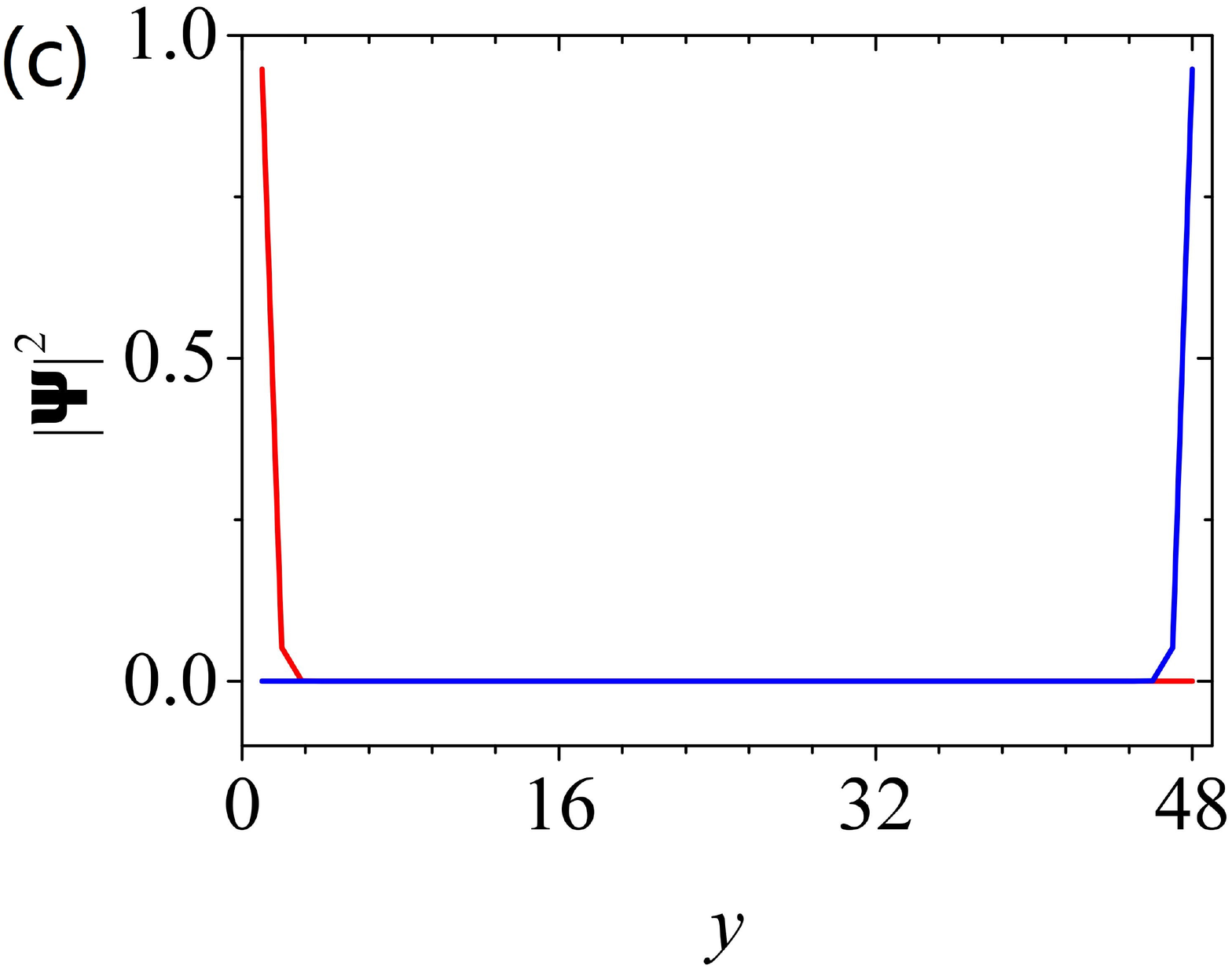}
\includegraphics[width=6.5cm]{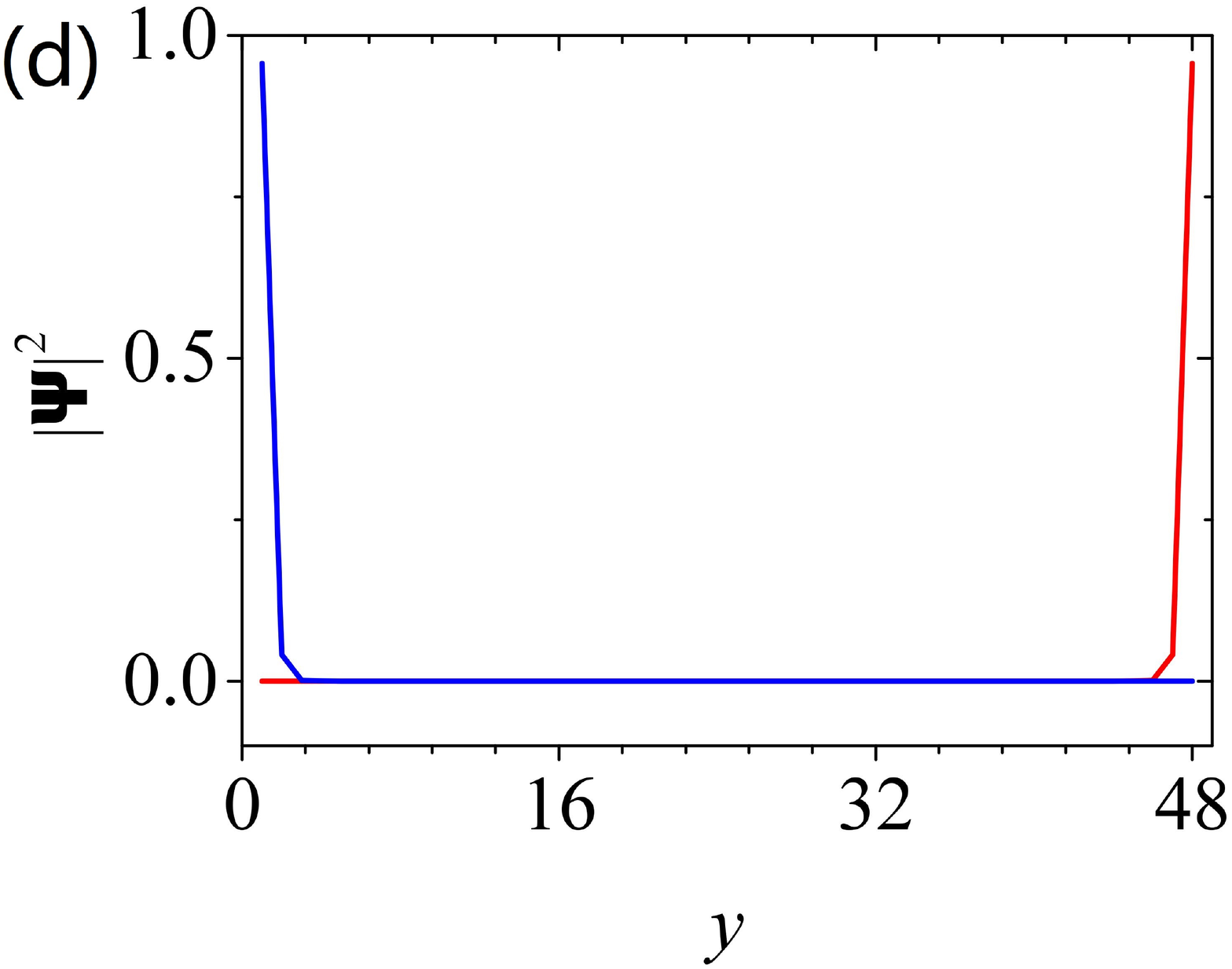}
\end{center}
\caption{(Color online)  (a) and (b) Energy spectrum of $d_2$ composite fermions in a cylinder geometry with periodic boundary conditions in the $x$-direction.
The calculations is done for $t_{1}=2\Delta_{1}=1$, $t_2=0.1$ with lattice size $48\times48$. The system has single chiral edge state on each edge for $\Delta_2=0$ in (a) and $\Delta_2=0.25$ in (b), but with the opposite chirality.
(c) and (d) are the real space wavefunction distribution of the edge states corresponding to (a) and (b). }
\label{fig:chiral}
\end{figure*}

The noninteracting Hamiltonian $H_0$ describes two components Haldane model with ESP at half-filling. 
According to the symmetry analysis in section \ref{symmetry}, the system falls into class $D$\cite{Schnyder} of topological superconductor (TSC). 
The topological invariant is given by the Chern number\cite{Thouless} $C_{\alpha}$, which denotes the Chern number of $d_{\alpha}$ composite fermions. 
We calculate the Chern numbers and find 
\begin{align}
C_{1} & =\mathrm{sign}\left(\tilde{t}_{2}\right)=\mathrm{sign}\left(t_{2}+\Delta_{2}\right)\nonumber \\
C_{2} & =\mathrm{sign}\left(\delta_{2}\right)=\mathrm{sign}\left(t_{2}-\Delta_{2}\right)
\end{align}
where $\mathrm{sign}(x)=\lim_{\epsilon\rightarrow0}\frac{x}{\sqrt{x^{2}+\epsilon^{2}}}$ is the sign function. 
We introduce the total Chern number and spin Chern number\cite{sheng2003phase,sheng2006quantum} as
\begin{align}
C & =C_{1}+C_{2}\nonumber \\
C_{\mathrm{spin}} & =C_{1}-C_{2}
\end{align}
which indicates the topological phase transition at $t_{2}=\pm\Delta_{2}$.
Accordingly the gap closes for $d_1$ ($d_2$) composite fermions at $t_{2}=-\Delta_{2}$ ($t_{2}=\Delta_{2}$).
The phase diagram of Haldane-BCS model is shown in FIG. \ref{fig:phase}.
For $\left|t_{2}\right|>\left|\Delta_{2}\right|$, the system is in the chiral TSC state with total Chern number $C=\pm2$ and spin Chern number $C_{\mathrm{spin}}=0$.
For $\left|t_{2}\right|<\left|\Delta_{2}\right|$, the system is in the helical TSC state with total Chern number $C=0$ and spin Chern number $C_{\mathrm{spin}}=\pm2$.
Along the critical lines $t_{2}=\pm\Delta_{2}$, one species of $d$ composite fermions is gapless and another species is in the chiral  TSC state with Chern number $C_{\alpha}=\pm1$.
The origin is a gapless multicritical point.

\begin{figure*}[tb]
\begin{center}
\includegraphics[width=6.5cm]{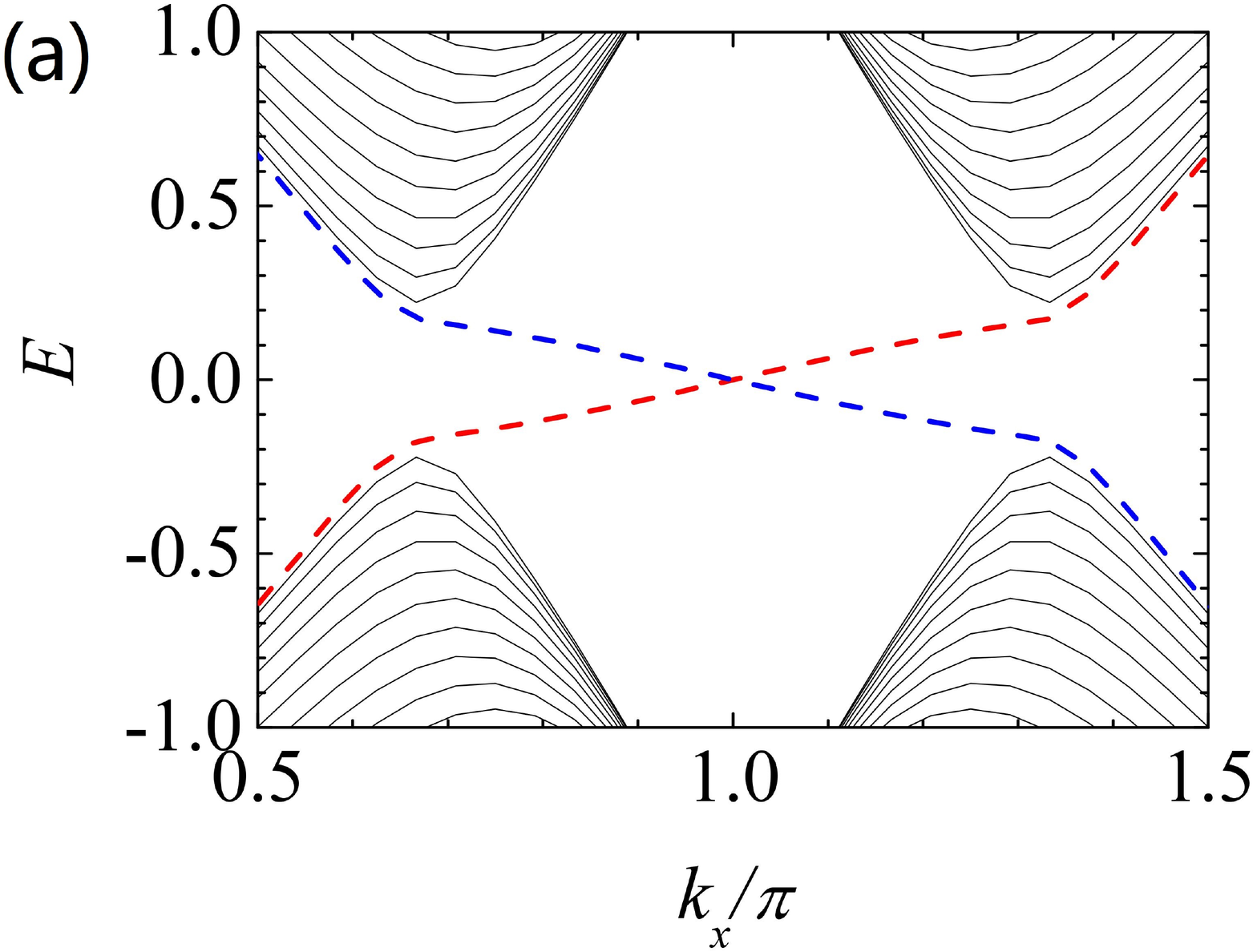}
\includegraphics[width=6.5cm]{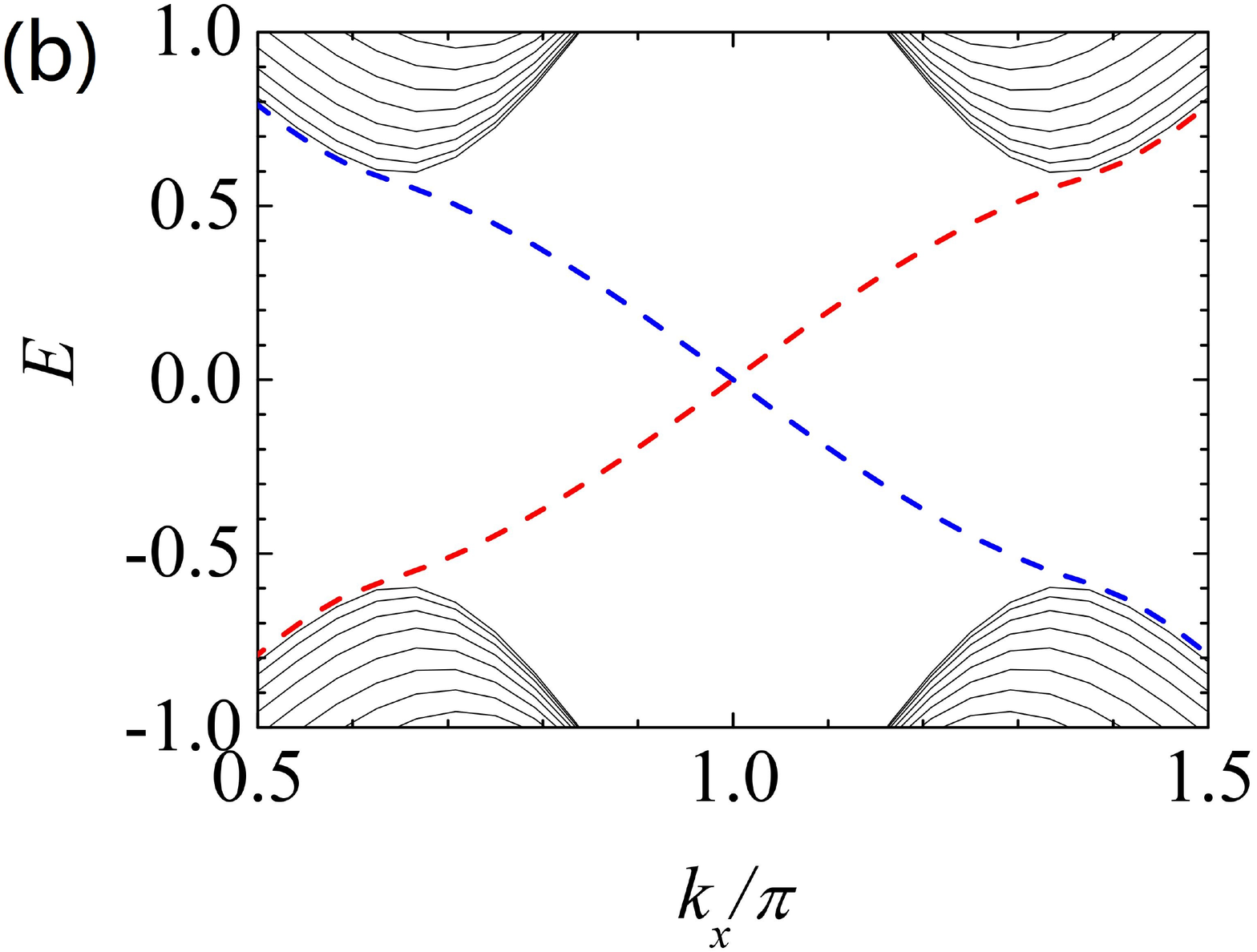}
\includegraphics[width=6.5cm]{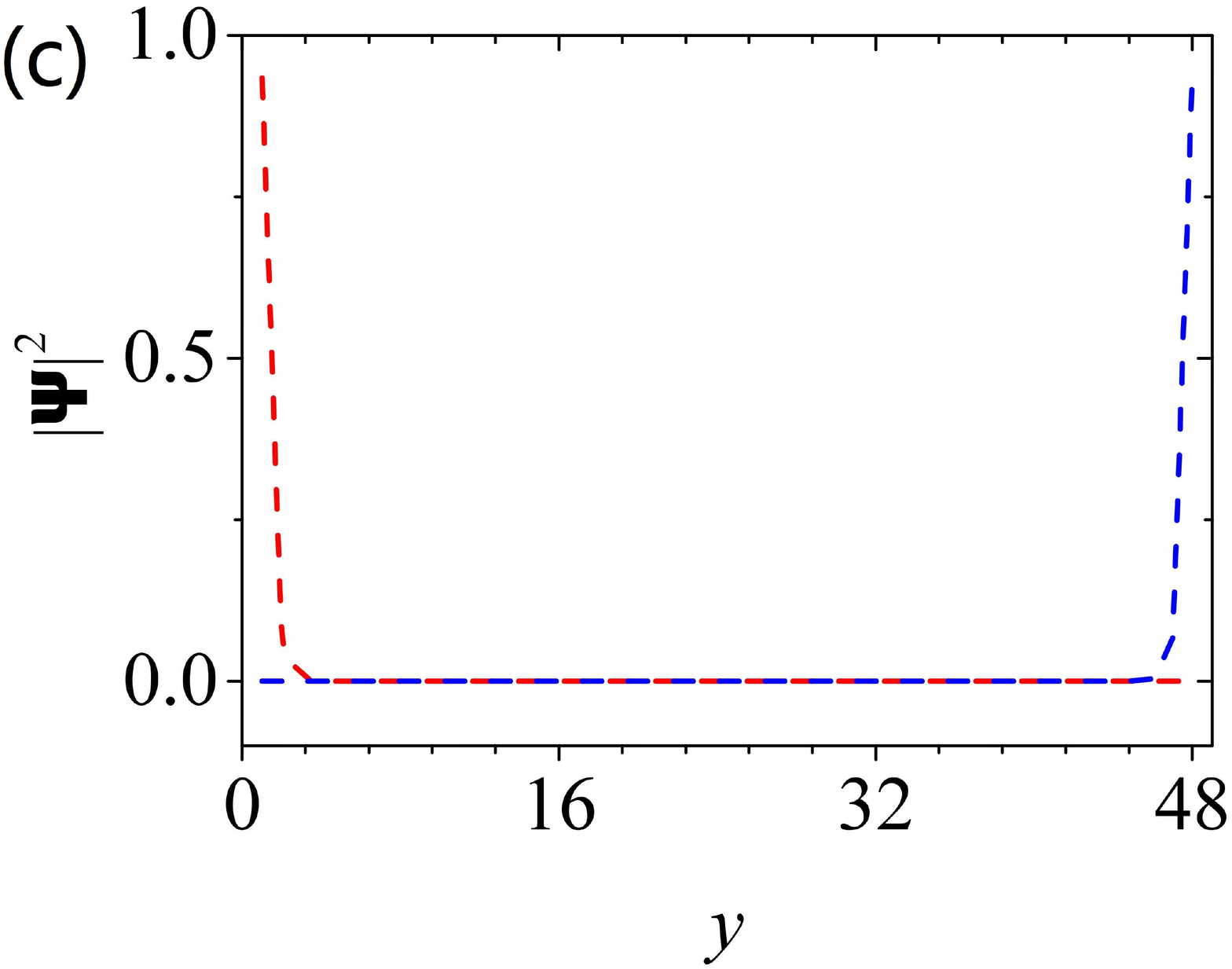}
\includegraphics[width=6.5cm]{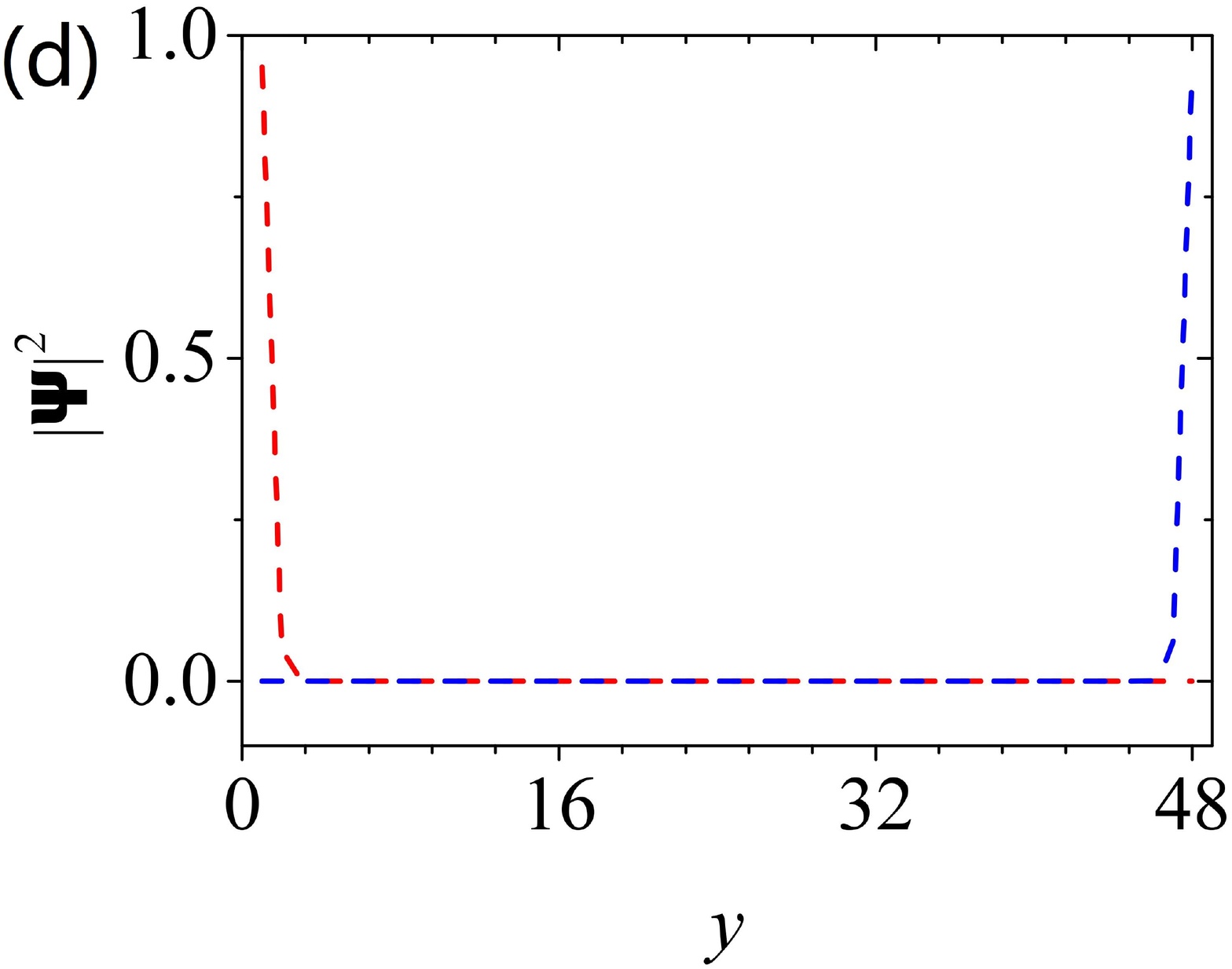}
\end{center}
\caption{(Color online)  (a) and (b) Energy spectrum of $d_1$ composite fermions in a cylinder geometry with periodic boundary conditions in the $x$-direction.
The calculations is done for $t_{1}=2\Delta_{1}=1$, $t_2=0.1$ with lattice size $48\times48$. The system has single chiral edge state on each edge for $\Delta_2=0$ in (a) and $\Delta_2=0.25$ in (b),with the same chirality.
(c) and (d) are the real space wavefunction distribution of the edge states corresponding to (a) and (b). }
\label{fig:non-chiral}
\end{figure*}

The topological phase transition can be understood via the bulk-edge correspondence.
Except along the critical lines, each species of $d$ composite fermions has nonzero Chern number, i.e. in the TSC state with single chiral edge state.
For $\left|t_{2}\right|>\left|\Delta_{2}\right|$, both edge states carry the same chirality and the system is in the chiral TSC state with two chiral edge states, 
which is consistent with total Chern number $C=\pm2$ and spin Chern number $C_{\mathrm{spin}}=0$. 
However for $\left|t_{2}\right|<\left|\Delta_{2}\right|$, two edge states have opposite chirality.
The system becomes a helical TSC state with total Chern number $C=0$ while spin Chern number $C_{\mathrm{spin}}=\pm2$.
We plot the energy spectrum of $d_2$ composite fermions with different sign of $\delta_2$ in Fig. \ref{fig:chiral} and find in both cases the system has single chiral edge state on each edge.
Due to sign change of $\delta_2$, the wavefunctions of the edge states localize on opposite edges, which indicates the chirality of the edge states is changed.
For comparision, we also show the energy spectrum and wavefunctions of $d_2$ composite fermions in Fig. \ref{fig:non-chiral} .
This is consistent with the sign change of Chern number of $d_2$ composite fermions.
Along the critical lines $t_{2}=\pm\Delta_{2}$, one chiral edge state merges into the bulk and the system becomes a gapless TSC state with single chiral edge state.
The topological phase transition can also be revealed by another dual mapping $\eta_{\lambda}\rightarrow\lambda\eta_{\bar{\lambda}}$ ($\gamma_{\lambda}\rightarrow\lambda\gamma_{\bar{\lambda}}$), where $\bar{\lambda}$ is the different sublattice of $\lambda$. 
The Hamiltonian has the dual symmetry with parameters changing as $t_{2}\leftrightarrow\Delta_{2}$ ($t_{2}\leftrightarrow -\Delta_{2}$).
The topological phase transition happens exactly along the self-dual lines $t_{2}=\pm\Delta_{2}$. 
Similar duality relating topological and trivial phases has been discovered in the interacting Kitaev chain\cite{Miao}.
If we employ the BdG formalism\cite{Qi} and use the Nambu spinor $\Psi_{k\alpha}^{\dagger}=\bigl(d_{k\alpha A}^{\dagger},d_{k\alpha B}^{\dagger},d_{-k\alpha A},d_{-k\alpha B}\bigr)$, the above analysis is still valid except the Chern numbers should  be multiplied by $2$ and each chiral edge state becomes two chiral Majorana edge states.

\section{Haldane-BCS-Hubbard model along symmetric lines}\label{interacting}
In this section, we analyze the Haldane-BCS-Hubbard model along the symmetric lines $\delta_1=\delta_2=0$.
In terms of $d$ composite fermions language, the $d_2$ fermions are completely localized (or form the completely flat bands in the band theory language\cite{Ezawa}).
The Hamiltonian effectively reduces to the Falicov-Kimball model with only one species of mobile $d$ composite fermions

As the total Hilbert space is divided into different sectors characterized by the sets of $\left\{ D_{r\lambda}\right\} $, we first determine the ground state sector.
Within each sector, the ground state energy is by summing all the negative energy levels.
The ground state sector is determined by the set of $\left\{ D_{r\lambda}\right\} $ with minimal ground state energy.
We traverse all the $2^N$ sectors numerically for small lattice size and find the ground state sectors are $\left\{ D_{r\lambda}=\pm\lambda\right\}$.
We also note the sectors $\left\{ D_{r\lambda}=\pm1\right\} $ have the maximal ground state energy.
For large lattice size, we randomly choose the sector $\left\{ D_{r\lambda}\right\} $ and find its ground state energy always falls between the sectors $\left\{ D_{r\lambda}=\pm\lambda\right\} $ and $\left\{ D_{r\lambda}=\pm1\right\} $.
The numerical details are given in the appendix \ref{numerical}.
Thus we conclude the ground state sectors are $\left\{ D_{r\lambda}=\pm\lambda\right\}$.
The ground states are uniform with two-fold degeneracy.

Within the ground state sectors, the Hamiltonian of the Haldane-BCS-Hubbard model along the symmetric lines reduces to
\begin{widetext}
\begin{align}
H_{s} & =-\frac{i\tilde{t}_{1}}{2}\sum_{r}\left(d_{r1A}^{\dagger}d_{r1B}^{\dagger}+d_{r1A}^{\dagger}d_{r+a_{1}1B}^{\dagger}+d_{r1A}^{\dagger}d_{r+a_{2}1B}^{\dagger}+h.c.\right)\nonumber \\
 & -\frac{i\tilde{t}_{2}}{2}\sum_{r\lambda}\lambda\left(d_{r1\lambda}^{\dagger}d_{r+a_{1}1\lambda}+d_{r1\lambda}^{\dagger}d_{r-a_{1}+a_{2}1\lambda}+d_{r1\lambda}^{\dagger}d_{r-a_{2}1\lambda}\right)+h.c.\nonumber \\
 & \pm\frac{U}{2}\sum_{r\lambda}\left(n_{r1\lambda}-\frac{1}{2}\right)
\end{align}
\end{widetext}
where $D_{r\lambda}=2\lambda\bigl(n_{r2\lambda}-\frac{1}{2}\bigr)=\pm\lambda$.
The $d_2$ composite fermions form the background $\mathbb{Z}_2$ charge fields.
Note the Hamiltonian is symmetric with respect to the Hubbard $U$ along the symmetric lines.
As the ground state sectors are translation invariant, we can perform the Fourier transformation and the Hamiltonian $H_s$ can also be written in the form of 
\begin{equation}
H_{s}=\sum_{k}\psi_{k1}^{\dagger}h_{s}\left(k\right)\psi_{k1}
\end{equation}
where 
\begin{equation}
h_{s}\left(k\right)=\vec{T}_{s}\left(k\right)\cdot\vec{\sigma}
\end{equation}
with $T_{s}^{x}=T_{1}^{x}$, $T_{s}^{y}=T_{1}^{y}$ and $T_{s}^{z}=T_{1}^{z}\pm\frac{U}{2}$.
The energy dispersion reads $E_{s}\left(k\right)=\pm\bigl|\vec{T}_{s}\left(k\right)\bigr|$, which is gapped except at $U=\pm3\sqrt{3}\tilde{t}_{2}$.
The quasiparticle excitations are the spin-$1/2$ $d_1$ composite fermions. 
Even with the Hubbard interactions, we can define the spectral Chern number in terms of these quasiparticles along the symmetric lines.
The spectral Chern number is given by 
\begin{equation}\label{spectrumchern}
C_s=\frac{1}{2}\left[\mathrm{sign}\left(3\sqrt{3}\tilde{t}_{2}-U\right)+\mathrm{sign}\left(3\sqrt{3}\tilde{t}_{2}+U\right)\right]
\end{equation}
Thus there is a topological phase transition at $U=\pm3\sqrt{3}\tilde{t}_{2}$ and the gap closes at this point accordingly.

This topological phase transition can be understood easily in terms of $d_1$ composite fermions.
Within each sector, the Hubbard interactions act as chemical potential terms.
For small $U$, the system is in the weak pairing region and topological.
While for large $U$, the system is in the strong pairing region and becomes topologically trivial\cite{Read, Qi_RMP}.
The topological phase transition is due to the competition between the NNNH terms and Hubbard interactions.
This mechanism is remarkably different from the Kane-Mele-BCS-Hubbard model studied in Ref. \onlinecite{Ezawa}, where an infinitesimal $U$ renders the topological SC state into trivial, because its topological SC state is protected by the TRS, and the Hubbard interaction always spontaneously breaks the TRS and mixes different spin components within each sector as
$Ti\gamma_{\uparrow\lambda}\gamma_{\downarrow\lambda}T^{-1}=-i\gamma_{\uparrow\lambda}\gamma_{\downarrow\lambda}$.
In the Haldane-BCS-Hubbard model along the symmetric lines, the topological phase transition happens at finite $U$, which clearly manifests the competition of topology and correlations.

\begin{figure}[tb]
\begin{center}
\includegraphics[width=8.0cm]{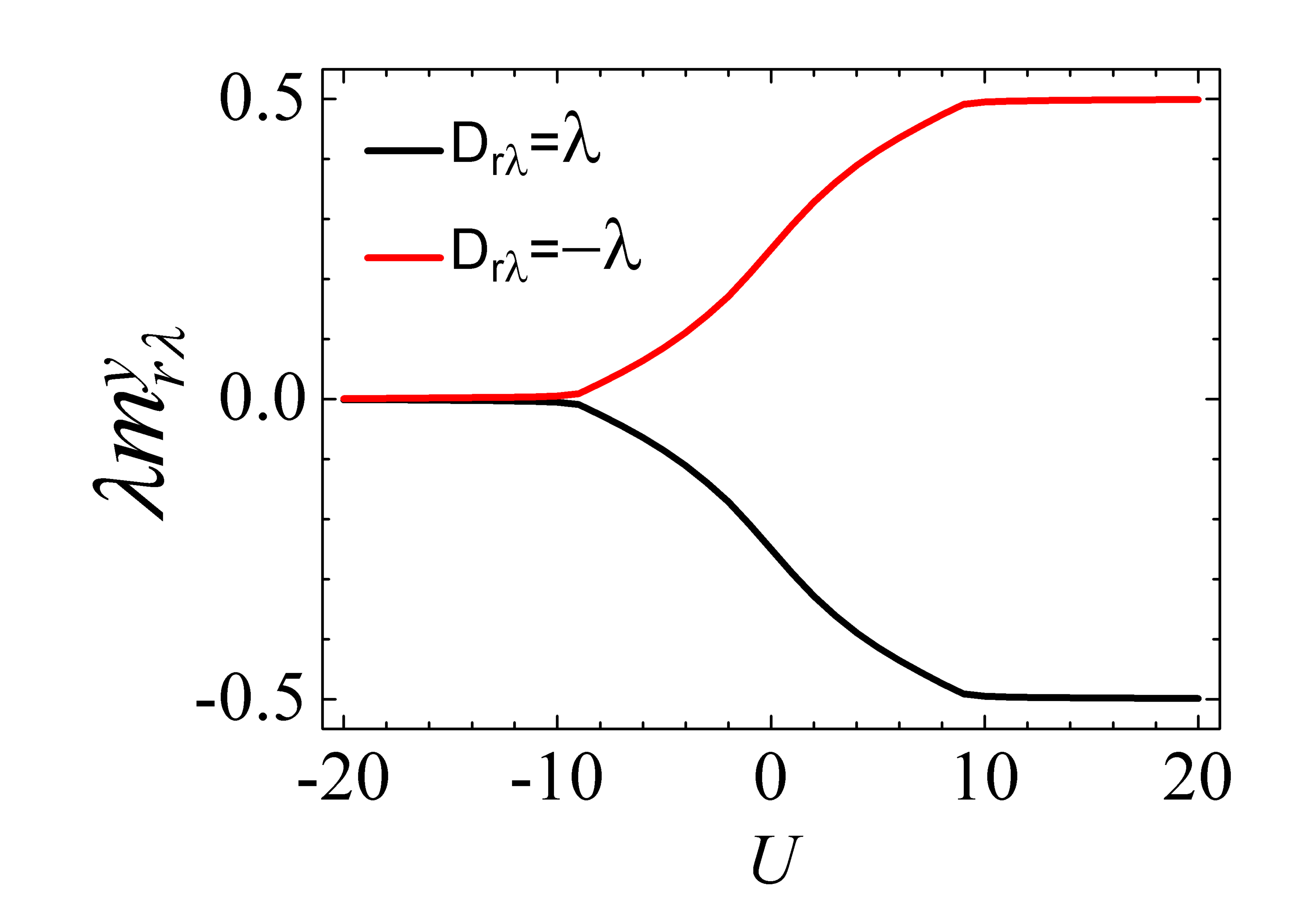}
\end{center}
\caption{(Color online) The transverse magnetism in the $y$-direction $\lambda m_{r\lambda}^{y}$ . The parameters are $\tilde{t}_{1}=1$ and $\tilde{t}_{2}=\sqrt{3}$.}
\label{fig:mag}
\end{figure}

We study the properties of ground states with the aid of symmetry analysis.
Even though the Hamiltonian does not have the inversion symmetry $I$, we note it has the combined symmetry $\tilde{I}$ of bond centered inversion $I$ plus gauge transformation $c_{rs\lambda}\rightarrow\sum_{s'}\left(i\sigma^{z}\right)_{ss'}c_{rs'\lambda}$ .
The two degenerate ground states are transformed to each other by the symmetry $\tilde{I}$.
Thus the ground states spontaneously break the $\mathbb{Z}_2$ symmetry $\tilde{I}$ for nonzero $U$.
We define the transverse magnetism in the $y$-direction as the order parameter\cite{Ng_2018}
\begin{equation}
m_{r\lambda}^{y}=\frac{1}{2}\bigl\langle c_{r+\lambda}^{\dagger}c_{r+\lambda}-c_{r-\lambda}^{\dagger}c_{r-\lambda}\bigr\rangle.
\end{equation}
We calculate the transverse magnetism via the operator identity
\begin{equation}
c_{r+\lambda}^{\dagger}c_{r+\lambda}-c_{r-\lambda}^{\dagger}c_{r-\lambda}=\lambda\left(d_{r1\lambda}^{\dagger}d_{r1\lambda}-d_{r2\lambda}^{\dagger}d_{r2\lambda}\right)
\end{equation}
For comparison, we find $\left\langle d_{r\alpha\lambda}^{\dagger}d_{r\alpha\lambda}\right\rangle =\frac{1}{2}$ in the noninteracting limit with generic hopping and pairing amplitudes, thus the ground state is nonmagnetic.
For nonzero $U$ along the symmetric lines, we have 
\begin{align}
\bigl\langle d_{r1\lambda}^{\dagger}d_{r1\lambda}\bigr\rangle & =\frac{1}{N}\sum_{k}\left(\frac{1}{2}\mp\frac{T_{s}^{z}\left(k\right)}{2E_{s}\left(k\right)}\right)\nonumber \\
\left\langle d_{r2\lambda}^{\dagger}d_{r2\lambda}\right\rangle  & =\frac{1}{2}+\frac{\lambda D_{r\lambda}}{2}
\end{align}
The order parameter is given by
\begin{equation}
m_{r\lambda}^{y}=\pm\frac{\lambda}{2}\left(\frac{1}{2}+\frac{1}{N}\sum_{k}\frac{T_{s}^{z}\left(k\right)}{2E_{s}\left(k\right)}\right)
\end{equation}
Thus the ground states have antiferromagnetic order for nonzero $U$.
The transverse magnetism shown in FIG. \ref{fig:mag} indicates the $\mathbb{Z}_2$ symmetry $\tilde{I}$ is spontaneously breaking.
In the limit $U\rightarrow\infty$, there is only one electron per site and the spin is fully polarized.
Accordingly we have $m_{r\lambda}^{y}\rightarrow\pm\frac{\lambda}{2}$.
In the limit $U\rightarrow - \infty$, each site is either empty or doubly occupied, thus we have $m_{r\lambda}^{y}\rightarrow 0$.
We give a remark on the nonzero magnetism at $U=0$ in FIG. \ref{fig:mag}.
For generic hopping and pairing amplitudes, the ground state is nonmagnetic for $U=0$. 
While along the symmetric lines, one species of composite fermion is completely flat. 
So it is possible to form the nonzero magnetism by the linear combination of these localized states and the curve is continuous at $U=0$. 
We note similar magnetic topological phase for small Hubbard $U$ has been found before\cite{yoshida2013topological}.

\section{Summary and discussions}\label{discussion}
In this paper, we study the Haldane-BCS-Hubbard model. We find this model can be solved exactly along the symmetric lines.
In the noninteracting limit, the Haldane-BCS model has topological phase transitions at the self-dual points. The topological phase transition are revealed by the bulk-edge correspondence.
Along the symmetric lines, we find the model reduces to the Falicov-Kimball model. 
There is an interaction induced topological phase transition due to the competition between NNNH terms and Hubbard interaction.
With nonzero Hubbard $U$, the ground states spontaneously break the $\mathbb{Z}_2$ symmetry and have staggered transverse magnetism in the $y$-direction.
 
The Haldane model has already been realized in the cold atoms system\cite{Jotzu}. Actually we can view our model as bilayer of Haldane models. The spin index $s$ can be viewed as the layer index with $s=\uparrow$ for the upper layer and $s=\downarrow$ for the bottom layer. The ESP and the on-site interaction Hubbard $U$ between two layers might be introduced in cold atom systems. Therefore, we expect the interaction induced topological phase transition can be observed in cold atom systems.

\begin{figure}[tb]
\begin{center}
\includegraphics[width=9.5cm]{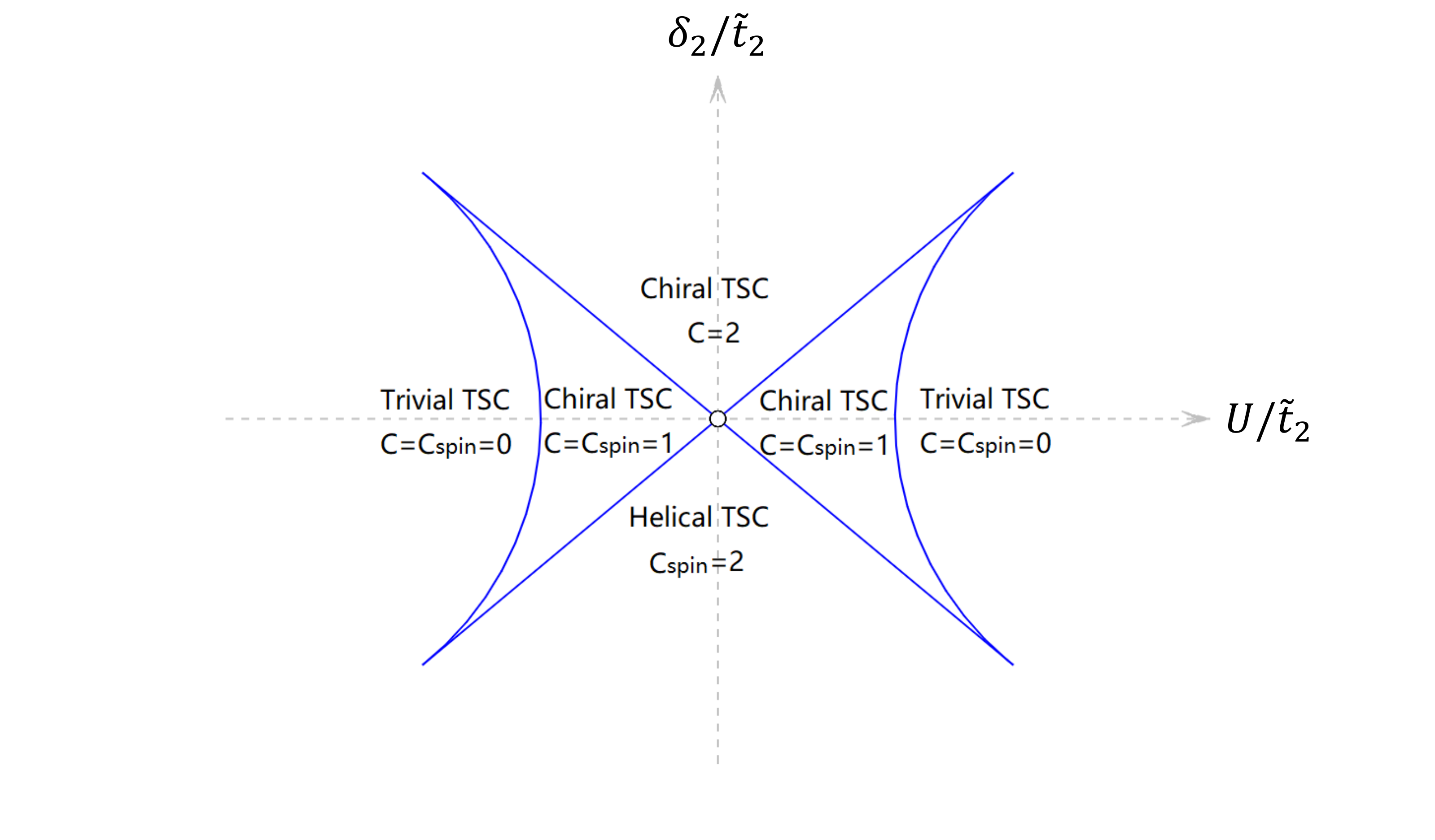}
\end{center}
\caption{(Color online) The sketched phase diagram away from the symmetric lines. The hopping and pairing parameters are all positive and $\delta_1\neq0$. The curves intersect the horizontal axis at $|U|=3\sqrt{3}\tilde{t}_{2}$.}
\label{fig:butterfly}
\end{figure}

Let us consider the topological characterization of the system with the Hubbard interaction for generic hopping and pairing parameters. The total Chern number $C$ and the spin Chern number $C_{\mathrm{spin}}$ can be defined with the many-body ground-state wavefunctions in the twisted boundary condition\cite{niu1985quantized,sheng2003phase}. However, the Chern numbers $C_{\alpha} (\alpha=1, 2)$ for each species of $d_{\alpha}$ composite fermions are not well-defined because the two species are entangled with each other. An exception is along the symmetric lines, where the $d_{2}$ fermions are completely localized, and the ground state wavefunction is given by $|\Psi\rangle=|\Psi_{1}(\{D_{r\lambda}\})\rangle\otimes |\Psi_{2}(\{D_{r\lambda}\})\rangle$, in which $\{D_{r\lambda}=\pm \lambda\}$ is selected by the interaction at the ground state, and $|\Psi_{1}(\{D_{r\lambda}\})\rangle$ and $|\Psi_{2}(\{D_{r\lambda}\})\rangle$ denote the wavefunctions of the $d_{1}$ and $d_{2}$ fermions for a given $\{D_{r\lambda}\}$ sector. The localized $d_{2}$ fermions do not respond to the twisted boundary condition, thus $C_{2}=0$. The $d_{1}$ fermions contribute the spectral Chern number given by Eq. (\ref{spectrumchern}). Therefore, the total Chern number and the spin Chern number are given by $C=C_{\mathrm{spin}}=C_{1}$ along the symmetric lines for nonzero $U$, which persist for nonzero $\delta_{1}$ and $\delta_{2}$ as long as the bulk gap is not closed due to the topological stability. For large $\delta_{1}$ and $\delta_{2}$, we expect topological phase transitions to phases that are adiabatically connected to the two gapped phases in the noninteracting case. Therefore, an interaction-induced new topological phase\cite{Qi,wang2015chiral} emerges for $0<|U|<3\sqrt{3}\tilde{t}_{2}$ intervening the two phases in the noninteracting case. We plot a schematic phase diagram in FIG.\ref{fig:butterfly}.

\section{Acknowledgement}
J.J.M. acknowledges the discussion with Tai-Kai Ng and Yi Zhou.
J.J.M. is supported by China Postdoctoral Science Foundation (Grant No.2017M620880) and the National Natural Science Foundation of China (Grant No.1184700424).
D.H.X. is supported by the National Natural Science Foundation
of China (Grant No. 11704106) and the Scientific Research Project of Education
Department of Hubei Province (Grant No. Q20171005). D.H.X. also acknowledges the support of the Chutian Scholars Program in Hubei Province.
L.Z. is supported by National Key R\&D Program of China (No. 2018YFA0305800) and National Natural Science Foundation of China (No. 11804337). Work at UCAS is also supported by Strategic Priority Research Program of CAS (No. XDB28000000), and Beijing Municipal Science \& Technology Commission (No. Z181100004218001).
F.C.Z. is supported by National Science Foundation of China (Grant No.11674278) and National Basic Research Program of China (No.2014CB921203).

\section{Note added}
During the preparation of this work, we learned a similar work on arXiv\cite{Ng_2019},
which also studied the extension of the BCS-Hubbard model\cite{Ng_2018} (now named as Majorana Falicov-Kimball Model\cite{prosko2017simple}) more thorough.

\appendix
\section{Numerical determination of ground state sectors $\left\{ D_{r\lambda}\right\} $ }\label{numerical}
\begin{figure}[tb]
\begin{center}
\includegraphics[width=9.0cm]{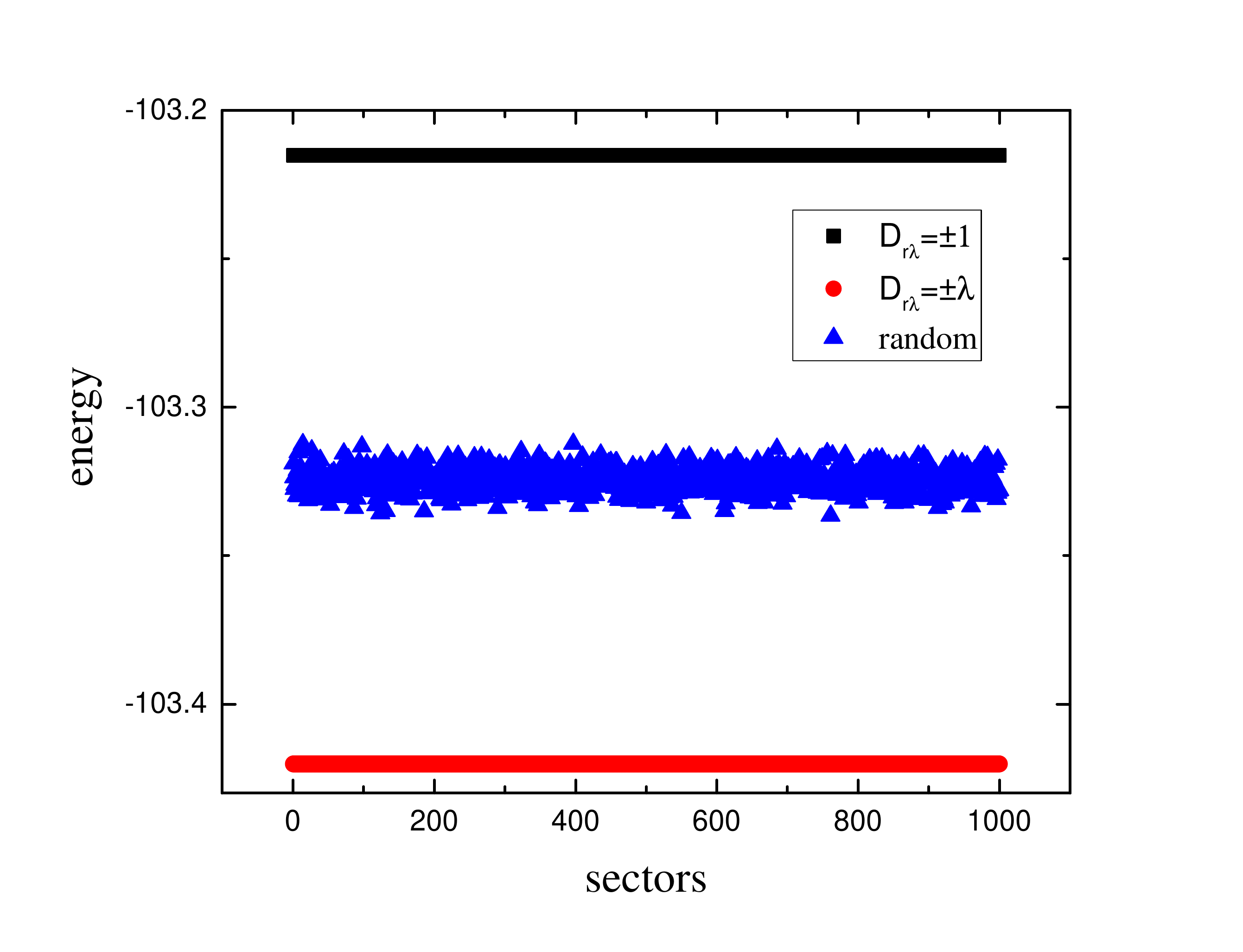}
\end{center}
\caption{(Color online)  ground state energy for various $\left\{ D_{r\lambda}\right\} $ sectors. Red points denote the sectors $\left\{ D_{r\lambda}=\pm\lambda\right\}$ with minimal ground state energy. Black points denote the sectors $\left\{ D_{r\lambda}=\pm1\right\}$ with maximal ground state energy. Blue points denote randomly chosen sectors, whose ground state energy falls between the sectors $\left\{ D_{r\lambda}=\pm\lambda\right\} $ and $\left\{ D_{r\lambda}=\pm1\right\}$. }
\label{fig:num}
\end{figure}

There are $2^N$ sectors characterized by the sets of $\left\{ D_{r\lambda}\right\}$, where $N$ is the number of total sites.
Up to $N=4\times4$, we can traverse all the $2^N$ sectors numerically on laptop in one minute.
By sorting all the sectors according to the ground state energy, we find the ground state sectors are $\left\{ D_{r\lambda}=\pm\lambda\right\}$ for arbitrary strength $U$.
We also note the sectors with the largest ground state energy are $\left\{ D_{r\lambda}=\pm1\right\}$.
For larger lattice size, the time and internal storage cost increase exponentially and it is impossible to traverse all the $2^N$ sectors numerically on laptop.
So we randomly choose the sector $\left\{ D_{r\lambda}\right\} $, i.e. the value of $D_{r\lambda}$ on each site is $1$ or $-1$ with equal weight, and calculate its ground state energy.
We find the ground state energy of randomly chosen sectors always falls between the sectors $\left\{ D_{r\lambda}=\pm\lambda\right\} $ and $\left\{ D_{r\lambda}=\pm1\right\} $.
For $N=16\times16$, we randomly choose $1000$ configurations and plot their ground state energy in FIG. \ref{fig:num}

\bibliographystyle{apsrev4-1}
\bibliography{ref}

\end{document}